\documentclass[journal]{IEEEtran}

\ifCLASSINFOpdf
\else
\fi

\usepackage{placeins}
\usepackage{times}
\usepackage{epsfig}
\usepackage{graphicx}
\usepackage{amsmath}
\usepackage{amssymb}
\usepackage{hyperref}
\usepackage{gensymb}
\usepackage{booktabs}
\usepackage{multirow}
\usepackage{verbatim}

\newcommand{\beq}{\begin{equation}}
\newcommand{\eeq}{\end{equation}}

\newcommand{\mr}[1]{\mathrm{#1}}

\usepackage{lineno,hyperref}
\usepackage{booktabs}
\usepackage{multirow}
\usepackage{enumitem}
\usepackage{amsmath,bm}
\usepackage{xcolor}
\modulolinenumbers[5]
\begin{document}
\title{Bare Demo of IEEEtran.cls\\ for IEEE Journals}
\author{Ahmad~Chaddad, Lama Hassan, Christian Desrosiers
\IEEEcompsocitemizethanks{\IEEEcompsocthanksitem A. Chaddad and Lama Hassan are with the School of Artificial intelligence, Guilin University of Electronic Technology. Guilin, Guanxgi, China. Ahmad Chaddad and Christian Desrosiers are with the The Laboratory for Imagery, Vision and Artificial Intelligence, Ecole de technologie Superieure, Montreal, Qc, Canada. 
\protect\\
E-mail: ahmad.chaddad@affiliate.mcgill.ca, ahmadchaddad@guet.edu.cn}
\thanks{Manuscript received June, 2021; revised August -, 2021.}}

%
%

\markboth{Journal of \LaTeX\ Class Files,~Vol.~14, No.~8, April~2021}%
{Shell \MakeLowercase{\textit{et al.}}: Bare Demo of IEEEtran.cls for IEEE Journals }
%

\title{Deep Radiomic Analysis for Predicting Coronavirus Disease 2019 in Computerized Tomography and X-ray Images}

\maketitle

\begin{abstract}
This paper proposes to encode the distribution of features learned from a convolutional neural network using a Gaussian Mixture Model. These parametric features, called GMM-CNN, are derived from chest computed tomography and X-ray scans of patients with Coronavirus Disease 2019. We use the proposed GMM-CNN features as input to a robust classifier based on random forests to differentiate between COVID-19 and other pneumonia cases. Our experiments assess the advantage of GMM-CNN features compared to standard CNN classification on test images. Using a random forest classifier (80\% samples for training; 20\% samples for testing), GMM-CNN features encoded with two mixture components provided a significantly better performance than standard CNN classification (p\,$<$\,0.05). Specifically, our method achieved an accuracy in the range of 96.00\,--\,96.70\% and an area under the ROC curve in the range of 99.29\,--\,99.45\%, with the best performance obtained by combining GMM-CNN features from both computed tomography and X-ray images. Our results suggest that the proposed GMM-CNN features could improve the prediction of COVID-19 in chest computed tomography and X-ray scans.
\end{abstract}
\begin{IEEEkeywords}
COVID-19, CNN, GMM, Radiomics.
\end{IEEEkeywords}
\section{Introduction}
Appeared in December 2019, the coronavirus disease 2019 (COVID–19) grew rapidly to become a world pandemic \cite{1}. Symptoms of infected patients, which range in intensity, include fever and fatigue with mainly dry cough and respiratory problems. While many patients show mild onset symptoms without fever, 2\,--\,4\% of infected subjects develop complications resulting in death \cite{3}. Various methods are used to detect COVID-19, including temperature measurement, molecular analysis (i.e., RT-PCR: reverse-transcription polymerase chain reaction), chest computed tomography (CT) scan and chest X-ray \cite{7}. Temperature is typically not considered as a definite measure for COVID-19 as it is a common marker for several other diseases. On the other hand, methods based on molecular examination (e.g., blood regimen, infection biomarkers, etc.) can generally validate the presence of COVID-19, however these methods tend to be costly, often take time for preparation and may cause side effects linked to illness. Moreover, studies have shown interstitial changes in early chest imaging of COVID-19 patients, specifically in the lateral area of the lung \cite{9}. In particular, ground-glass opacity (GGO) has been associated with extreme infection cases. Although RT-PCR test is considered as first procedure, imaging offers an efficient approach to assess the presence of COVID-19. In some cases, patients with negative first RT-PCR test and positive chest CT were found to test positive in a second RT-PCR test performed a few days later \cite{11}. The low sensitivity of RT-PCR tests has also been observed in screening COVID-19 patients \cite{12}. 

\begin{figure}[t!]
 \centering
 \includegraphics[width=0.48\textwidth]{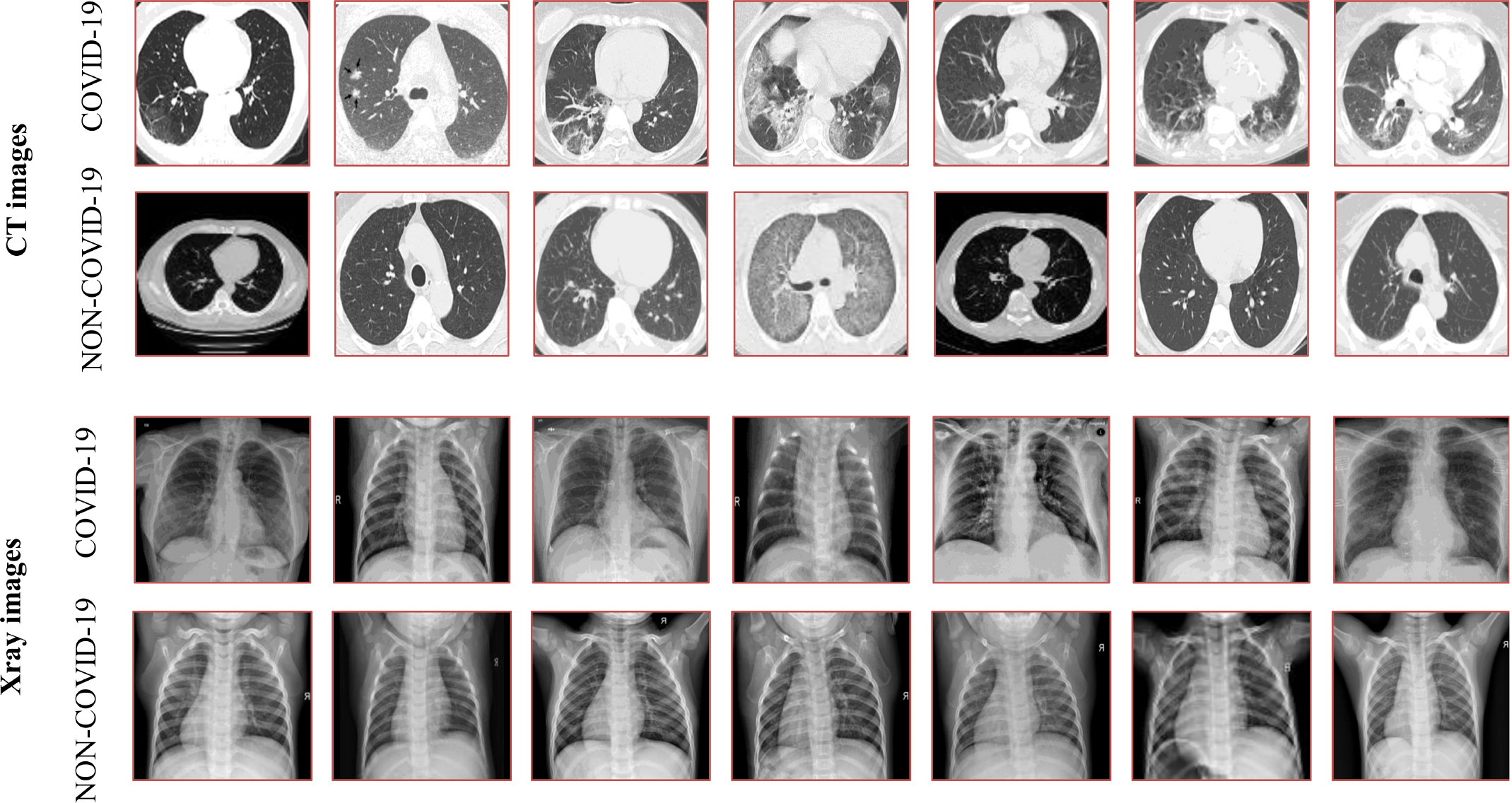}
 \caption{CT and X-ray image examples in different patient groups. From top (CT images) to bottom (x-ray images): COVID-19 vs. other pneumonia (NON-COVID-19). }
 \label{fig1}
\end{figure}

Image-based techniques offer a non-invasive alternative to identify the presence of COVID-19. Previous works have demonstrated the potential of using predictive models with CT imaging to diagnose COVID-19 \cite{14}. Chest CT has shown a high sensitivity for COVID-19 diagnosis \cite{12}, and X-ray images provide visual indexes related with such infection \cite{76}. Moreover, a recent study suggests that clinicians could rely on positive X-ray images showing the low or high extent of pneumonia, whereas intermediate extent cases identified by X-ray should be complemented by CT for optimal risk assessment \cite{77}. On the other hand, other studies have found that COVID-19 shared similar imaging features with viral pneumonia \cite{17}. Fig. \ref{fig1} shows CT and X-ray images of COVID-19 and NON-COVID-19 patients. We can observe GGO patterns with multifocal, bilateral, and peripheral lesions in CT images. Such lesions are commonly located in the inferior lobe \cite{81}. Similarly, X-ray images may also present GGO and consolidation regions. As COVID-19 cases share similar patterns as other pneumonia cases, differentiating between these two types of infection may require expert radiologists. Considering the large number of patients and the limited availability of radiologists in a pandemic, automatic detection techniques could increase the rate and accuracy of early diagnosis. 

Predictive models based on radiomics have led to a paradigm shift in analyzing complex medical data \cite{tamal2021integrated}. This approach, which extracts high-throughput features from digital medical images and uses them for various clinical prediction tasks, has had a high impact in medical image analysis and computer-aided diagnosis (CAD) \cite{21}. For example, CT radiomic models have been proposed for predicting pneumonia in patients with COVID-19 \cite{84} and assisting clinical decision-making \cite{23}. Recently, deep learning algorithms have also been applied to CT images for the automated detection of COVID-19 \cite{26} and to classify bacterial from viral pneumonia in pediatric chest radiographs \cite{27}. The main advantage of using convolutional neural networks (CNNs), compared to the conventional radiomic approach, is that features are learned directly from the data \cite{85}. However, this data-driven strategy is prone to overfitting when few labeled examples are available, leading to a poor generalization of new data \cite{bejani2021systematic}. To overcome this problem, the work in \cite{28} proposed a radiomic model that uses the conditional entropy of CNN features as input to a separate classifier. In \cite{82}, a patch-based CNN model with a limited number of trainable parameters was used to predict COVID-19 from CT images. Another strategy to alleviate the problem of overfitting in image classification is transfer learning \cite{30}. This strategy, which reuses the features of a neural network trained on a related task, has been used for COVID-19 detection in CT and X-ray imaging \cite{39}. 

To overcome the limitations of standard radiomics and deep learning methods for the image-based detection of COVID-19, we propose a novel approach that models the distribution of features in a pretrained CNN with a Gaussian mixture model (GMM). We hypothesize that feature maps in deep CNNs encode various scales of image texture that represent the heterogeneity of tissues. The main contributions of our work are as follows: 
\begin{itemize}\setlength\itemsep{.5em} 
\item We propose a new radiomic signature that encodes pretrained CNN features using a GMM. This signature captures characteristics of tissue heterogeneity that can effectively predict COVID-19 in CT and X-ray images;
\item We study the value for COVID-19 prediction of GMM-CNN descriptors from two different deep CNN architectures, DarkNet and ResNet; 
\item We show the effectiveness of the proposed features as input to a random forest classifier for discriminating between COVID-19 and other viral pneumonia. Our method achieves state-of-art performance in terms of accuracy on a publicly available dataset.
\end{itemize} 
The rest of this paper is structured as follows. Section II presents related works on radiomics and GMM applied to image analysis. Section III then describes the data used in this analysis and the proposed deep radiomic pipeline based on CNN and GMM. Afterwards, we present experimental results in Section IV and discuss our main findings in Section V. Last, Section VI summarizes the key contributions and results of our work.

\section{Related works}

Radiomics, which uses features extracted from medical image data to build prediction models, has become an important research subject in medical imaging and diagnostic radiology \cite{42}. However, since they use a large set of imaging features, radiomics models often suffer from overfitting which leads to a poor generalization on test data \cite{45}. Additionally, because these features are hand-crafted, they may not be optimal for a given prediction task. Deep learning models like CNNs aim to solve the latter problem by learning features directly from training images. Such models have been used successfully for disease diagnosis and treatment planning \cite{46}. Despite their soaring popularity for natural image analysis, CNNs have had a more limited success in clinical applications, due to the lack of labeled data for training \cite{huang2021artificial}. To overcome these limitations, recent studies proposed using a compact yet discriminative set of CNN features capturing image textures \cite{28}. For example, descriptors based on the entropy of CNN feature maps have been successfully applied to predict Alzheimer's disease in \cite{chaddad2018deep}. A texture analysis based on Gray-level co-occurrence matrices (GLCM) of CNN feature maps was also proposed to predict the survival of patients with recurrent glioblastoma (GBM) \cite{47}. 

The Gaussian Mixture Model (GMM) assumes that observed samples are generated from a combination of $k$ Gaussian components \cite{49}. The parameters of this model are the weight, mean vector and covariance matrix of each component, and are typically estimated from data using the Expectation-Maximization (EM) algorithm \cite{48}. The number of components can be determined on a validation set, in a supervised setting like ours, or using a metric like Bayesian information criterion that accounts for mode complexity and goodness of fit \cite{51}. In \cite{52}, a GMM with $k\!=\!3$ components is used to detect regions corresponding to brain tumors. Recently, several studies have suggested using a hybrid model based on GMMs and CNNs. For example, a GMM is used in conjunction with a CNN in \cite{53} for the unsupervised segmentation of histopathological images. Specifically, the CNN output is used to iteratively refine the color distribution of different tissue classes. GMM is often applied to extract discriminative features from the last layer of a CNN. Because this deep GMM model is trained end-to-end, it is often subject to overfitting when labeled data is limited \cite{53}. A GMM-CNN hybrid model was proposed in \cite{54} for brief statement speaker recognition, helping reduce the error rate compared to a standard approach. Likewise, the work in \cite{55} uses a set of 1024 features from a fully connected layer, in combination with other feature extraction techniques (i.e., local binary pattern, histogram of oriented gradients, dense scale invariant feature transform, etc.) and a Hidden Markov Model (HMM), to recognize house numbers in street view images. For medical image classification, a GMM was applied to extract features of pancreatic cancer images that were then used as input to a CNN classifier \cite{56}. A recent study used a GMM to encode features of a 3D CNN for predicting the survival of pancreatic cancer patients \cite{57}. However, it considered a shallow CNN with only 2 convolutional layers, resulting in a very limited set of features. In contrast, the current work explores the benefit of features from deep CNN architectures (i.e., DarkNet and ResNet) for detecting COVID-19 in 2D chest CT and X-ray images. 

\section{Methodology}

\begin{figure*}[t!]
 \centering
 \includegraphics[width=0.99\textwidth]{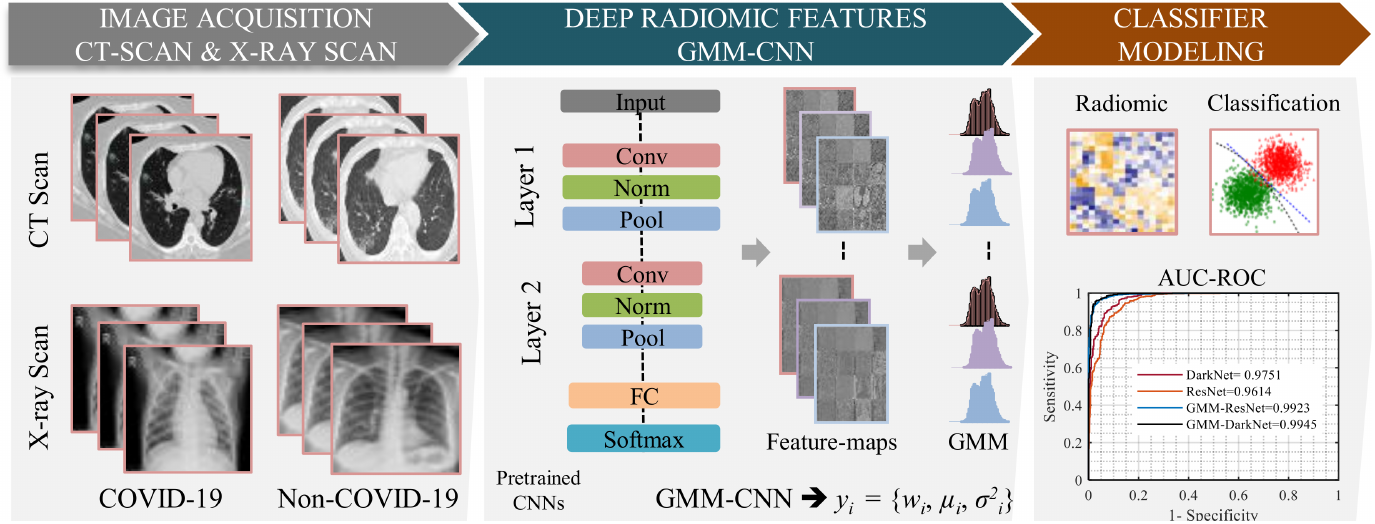}
 \caption{The proposed radiomic model for detecting COVID-19 in CT and Xray images: 1) CT-Xray scans are first acquired for COVID-19 patients; 2) The parameters of a GMM on each CNN feature map are extracted and used as multiscale texture descriptors. 3) The proposed descriptors are fed to a RF classifier for the final prediction.}
 \label{fig2}
\end{figure*}

The proposed radiomic model uses GMMs to extract multiscale texture information from the feature maps of a deep CNN. Specifically, we encode the distribution of CNN features from different layers using the GMM parameters ($\omega, \mu, \sigma^2$). This compact representation, called GMM-CNN, is then given as input to a random forest (RF) classifier to discriminate between COVID-19 and NON-COVID-19 cases. Fig. \ref{fig2} shows the flowchart of our radiomic model. 

\subsection{Patients and Data Acquisition}

A total of 5,254 axial CT (COVID-19 = 2,628; NON-COVID-19 = 2,626) and 8,084 X-ray (COVID-19 = 4,042; NON-COVID-19 = 4,042) images were obtained from a public dataset\footnote{https://data.mendeley.com/datasets/8h65ywd2jr/2}. These images were collected from multiple sites for the worldwide development and validation of AI applications to fight COVID-19. NON-COVID-19 cases correspond to other respiratory infections like viral and bacterial pneumonia. All images were previously deidentified by clinicians, and no institutional review board or Health Insurance Portability and Accountability Act approvals were preconditioned for our study. As pre-processing, images were resized, greyscaled, and normalized to the [0; 255] range with a size of 256$\times$256 pixels.

\subsection{Proposed radiomic descriptors}

Deep CNNs are widely used in medical image analysis, showing impressive performance for tasks like image classification and segmentation, particularly when large datasets are available \cite{58}. In general, a CNN is composed of a repeated stack of convolution and pooling layers, accompanied by one or more fully-connected layers and an output layer (e.g., softmax) that transforms logits into class probabilities. The convolutional filters and fully-connected layer weights are updated iteratively during training using the back propagation algorithm. Pre-trained CNNs can be used to reduce processing, as well as to improve training with limited data based on the principle of transfer learning. However, because of their very high learning capacity, deep CNNs can easily overfit the training data when few examples are available.

Unlike for natural image classification, where classes are recognized from salient objects in the image, detecting infections like COVID-19 in CT or X-ray images involves finding subtle and often local changes in image texture, which are captured by several layers of a CNN. Based on this idea, we generate texture descriptors by modeling the distribution of CNN features at different layers of a CNN using GMMs. Two deep CNN architectures are considered: ResNet50 \cite{59} and DarkNet-19 \cite{60}. The former is based on residual blocks that reuse features from previous layers via skip connections \cite{62}, thereby improving gradient flow. On the other hand, DarkNet is a standard feed-forward network used as classification module in the YOLOv4 object detection network \cite{60}. Both networks were pretrained on the ImageNet dataset, which comprises over 14 million natural images belonging to around 20,000 categories. We considered  stochastic gradient descent with momentum (SGDM) to update the network parameters with the batch size, learning rate and epochs number of 10, 1$\times 10^{-4}$, 10, respectively. 

\begin{figure*}[t!]
 \centering
 \includegraphics[width=0.99\textwidth]{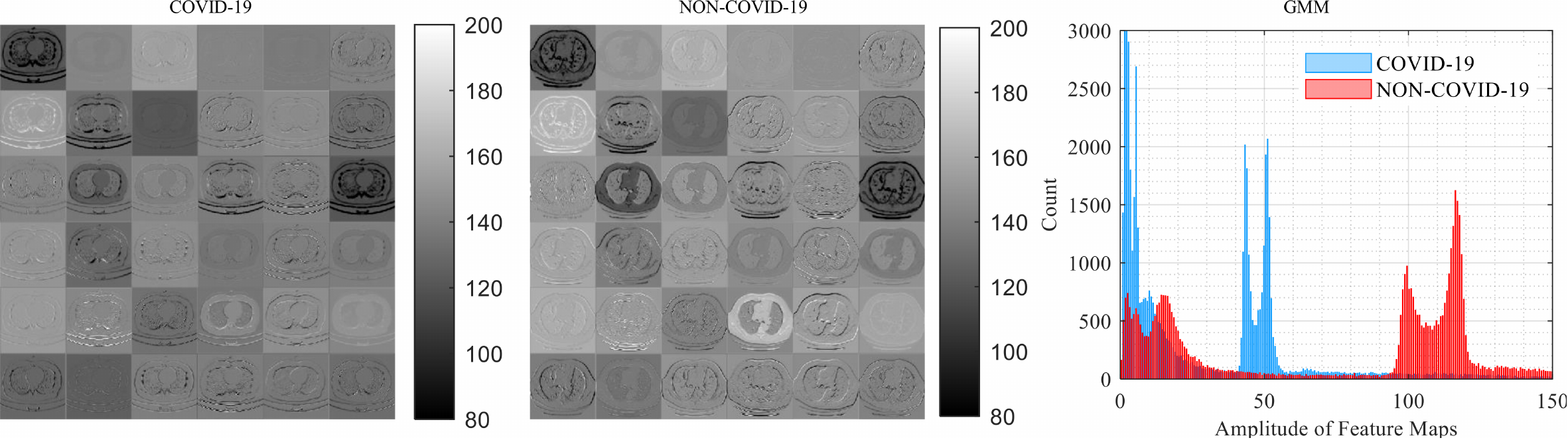}
 \caption{Example of feature maps from the first convolutional layer, i.e., $conv2d_1$, for COVID-19 and NON-COVID-19 cases (left) with their histograms (right).}
 \label{fig3}
\end{figure*}

Fig. \ref{fig3} shows examples of feature maps learned by DarkNet for COVID-19 and NON-COVID-19 cases. Although it is hard to see differences while inspecting these images, those differences become clearer when looking at the histograms of values in a feature map. Hence, feature distributions for COVID-19 and NON-COVID-19 have a similar number of modes, however the position of these modes differs. This observation motivates our use of GMM component parameters as a compact yet discriminative representation. Formally, let $\mathbf{z}_{\ell,i}$ be the $i$-th feature map of layer $\ell$ in the network. We model the distribution $p(Z_{\ell,i})$ as a weighted sum of $k$ Gaussian components, i.e.,
\beq
 p(Z_{\ell,i}) \, =\, \sum_{j\,=\,1}^k \omega^{(j)}_{\ell,i} \, \mathcal{G}\big(Z_{\ell,i}; \, \mu^{(j)}_{\ell,i}, (\sigma^{(j)}_{\ell,i})^2\big),
\eeq
where $\omega^{(j)}$ is the mixture weight of the $j$-th component, and $\mathcal{G}$ is the Gaussian probability distribution function with mean $\mu^{(j)}$ and variance $(\sigma^{(j)})^2$ defined as
\beq
\mathcal{G}(x; \mu,\sigma^2) \ = \ \frac{1}{\sigma \sqrt{2\pi}} \exp\!\Big(\!\!-\!\frac{1}{2\sigma^2} (x-\mu)^2\Big).
\eeq
Using this model, we represent feature map $\mathbf{z}_{\ell,i}$ by a row vector with $3\!\times\!k$ elements:
\beq
 \mathbf{y}_{\ell,i} \ = \ \big[\mu^{(1)}_{l,i}, \sigma^{(1)}_{l,i}, \omega^{(1)}_{l,i}, \, \ldots \, , \mu^{(k)}_{l,i}, \sigma^{(k)}_{l,i}, \omega^{(k)}_{l,i}\big].
\eeq
Let $m_\ell$ the number of feature maps in layer $\ell \in \{1, \ldots, n\}$. Considering feature maps in each convolutional layer and in the first fully-connected layer, we thus get a radiomic descriptor $\mathbf{y}_{\mr{\textsc{gmm-cnn}}}$ with a total of $3\!\times\!k\!\times \sum_{\ell=1}^n m_{\ell}$ features, where $n$ is the total number of layer and $m_\ell$ is the number of feature maps in layer $\ell$:
\beq
 \mathbf{y}_{\mr{\textsc{gmm-cnn}}} \ = \ [\mathbf{y}_{1,1} \, \cdots \, \mathbf{y}_{1,m_1} \ \cdots \ \mathbf{y}_{n,1} \, \cdots \, \mathbf{y}_{n,m_n}].
\eeq

We compare our GMM-CNN radiomic model with features obtained by running principal component analysis (PCA) on the ImageNet dataset. In this baseline, we first compute the covariance matrix of features maps in each network layer, using training examples of ImageNet. The main eigenvectors of this matrix (i.e., the principal components) are then used to project the feature maps of CT slices or X-ray images on a linear subspace with maximum variance. In this work, we used $\mr{PC}\!=\!3$ principal components, as it represented over $98\%$ of variance in most feature maps, giving a vector $\mathbf{y}_{\mr{\textsc{pca}}}$ with $3\!\times\!\sum_{\ell=1}^n m_{\ell}$ features for each image.
 
\subsection{Classification and performance metrics}
Once computed, the GMM-CNN or PCA features are then used as input to a RF classifier for discriminating between COVID-19 and NON-COVID-19 images. Although several other classifiers could be used for this task, e.g., support vector machines (SVMs) \cite{80}, we chose RF as it has few hyper-parameters to tune (i.e., number of trees, the maximum tree depth, etc.) and has a built-in feature selection process that provides interpretability \cite{65}. In addition, RFs can reduce the error due to data variance by exploiting a combination of decision tree bagging and random feature subspace selection \cite{66}. To tune the hyper-parameters of this model, we performed a grid search on the validation set and selected the following setting: \emph{Num. decision trees}\,=\,500, \emph{Max. tree depth}\,=\,15.

We evaluate the performance of using the proposed features with a RF classifier, and compare this approach to fine-tuning the network on the COVID-19 prediction task. Toward this goal, we divide the data in two subsets with different subjects, a training set with 80\% of examples and a test set with 20\% of examples. Additionally, 20\% of examples in the training set is set aside for validation. Performance is measured based on Accuracy, Sensitivity and Specificity:
\beq
\mr{\emph{Accuracy}}  \, = \, \frac{TP + TN}{N} \times 100 \\
\eeq
\beq
\mr{\emph{Sensitivity}}  \, = \, \frac{TP}{TP + FN} \times 100 \\
\eeq
\beq
\mr{\emph{Specificity}}  \, = \, \frac{TN}{TN + FP} \times 100 \\
\eeq

where $TP$ ($TN$, resp.) is the number of correctly predicted COVID-19 (NON-COVID-19) examples, $FP$ ($FN$, resp.) the number of examples incorrectly predicted as COVID-19 (NON-COVID-19, resp.), and $N$ the total number of examples. The latter two metrics are used to derive the area under the curve (AUC) of the receiver operator characteristic (ROC) curve, which plots the \emph{True positive rate} (\emph{Sensitivity}) versus \emph{False positive rate} (1-\emph{Specificity}) at different decision thresholds. The statistical significance of performance differences between the tested models is assessed with a chi-square test \cite{67}. All processing and analysis steps were performed using Matlab's Deep Learning \cite{paluszek2020practical}, Statistics, and Machine Learning Toolbox \cite{paluszek2016matlab}. 

\begin{table}[ht!]
\centering
\caption{Performance of our GMM-CNN model for $k \in \{2, 3, 4\}$ GMM components on 5 different cross-validation folds.
}\label{table1}
\setlength{\tabcolsep}{4pt}
\renewcommand{\arraystretch}{1.1}
\begin{small}
\begin{tabular}{cccccccc}
\toprule
& \multicolumn{2}{c}{$k=2$} & \multicolumn{2}{c}{$k=3$} & \multicolumn{2}{c}{$k=4$}\\
\cmidrule(l{5pt}r{5pt}){2-3}\cmidrule(l{5pt}r{5pt}){4-5}\cmidrule(l{5pt}r{5pt}){6-7}
Fold & Accuracy & AUC & Accuracy & AUC & Accuracy & AUC \\
\midrule
1 & 96.04& 98.99& 96.60& 98.78& 90.96& 96.42 \\
2 & 96.84& 99.25& 95.98& 99.04& 89.15& 96.66 \\
3& 96.66& 99.12& 95.73& 99.17& 90.96& 97.10 \\
4& 96.72& 99.28& 96.47& 98.85& 89.63& 96.26 \\
5 & 97.09& 99.29& 97.03& 99.10& 89.15& 96.65 \\
\midrule
Avg. & 96.67& 99.18& 96.36& 98.98& 89.97& 96.61 \\
\bottomrule
\end{tabular}
\end{small}
\end{table}

\begin{table}[ht!]
\centering
\caption{p-values of chi-square tests comparing the predictive models on CT and X-ray images.
}\label{table2}
\setlength{\tabcolsep}{5pt}
\renewcommand{\arraystretch}{1.1}
\begin{small}
\begin{tabular}{llccc}
\toprule
Images & Models & GMM-DarkNet\,+\,RF & GMM-ResNet\,+\,RF \\
\midrule
CT& DarkNet& 0.003& 0.01 \\
CT& ResNet& 0.001& 0.005 \\
X-ray& DarkNet& 0.02& 0.01 \\
X-ray & ResNet& 0.01& 0.03 \\
\bottomrule
\end{tabular}
\end{small}
\end{table}

\section{Experimental results}


We first test the proposed GMM-CNN model using combined feature maps from the DarkNet and ResNet networks, and both CT and X-ray images. Different numbers of GMM components $k\!\in\!\{2,3,4\}$ are considered for this analysis. To have a better estimate of performance in this setting, we employ a 5-fold cross-validation (CV). In this strategy, available samples are randomly partitioned into 5 even-sized subsets and, for each fold, a different subset is used as test set while remaining samples are considered as training set (20\% of these samples are set aside for validation). Table \ref{table1} reports the accuracy and AUC of the RF classifier using GMM-CNN features, for each CV fold. Overall, our GMM-CNN model using features from pretrained DarkNet and ResNet networks yields a high performance for different values of $k$, with an accuracy (AUC-ROC) range of 89.97\,--\,96.67\% (99.18\,--\,96.61\%). The highest accuracy (AUC-ROC) of 96.67\% (99.18\%) is achieved for $k\!=\!2$ GMM components, showing the discriminative ability of our compact set of features. Although not reported in the table, we found that using a larger number of components $k\!\in\!\{5, \ldots, 10\}$ further decreased performance, giving an average accuracy (AUC-ROC) near 85\% (90\%) for $k\!=\!10$. This suggests that components beyond $k\!=\!2$ or $k\!=\!3$ encode image-specific information and noise which leads to overfitting when learned. 

\begin{figure}[ht!]
\centering
\begin{small}
\includegraphics[width=0.49\textwidth]{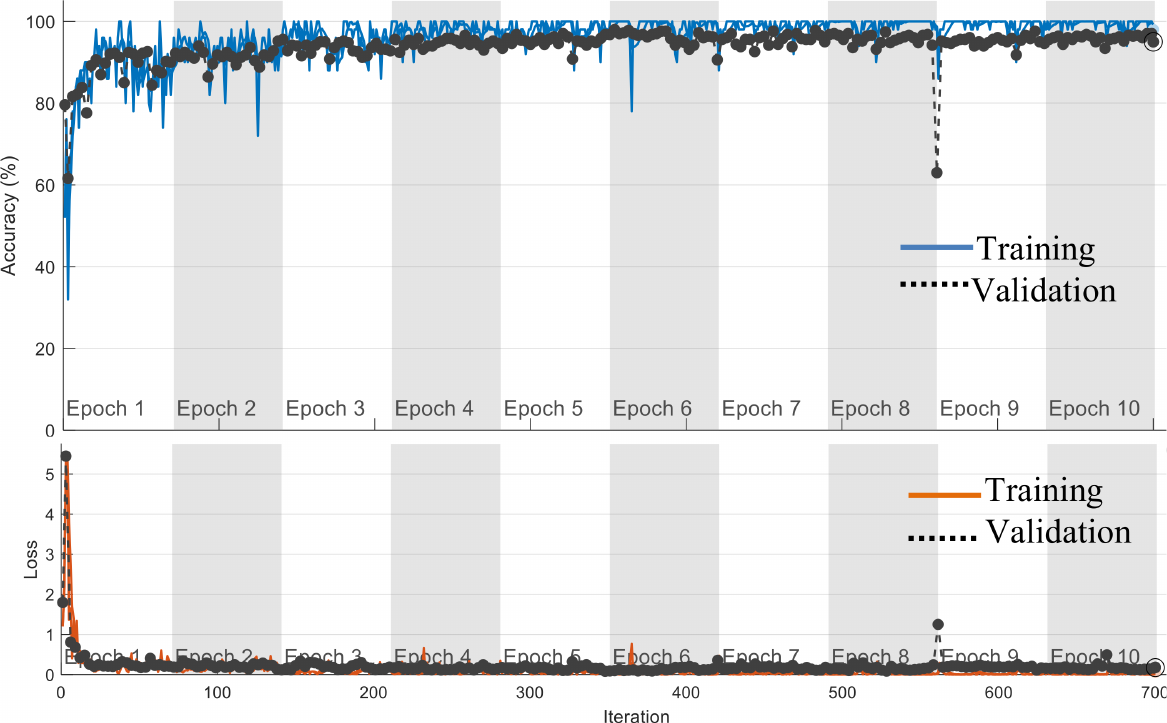} \\
(a) \\
\vspace{0.25cm}
\includegraphics[width=0.49\textwidth]{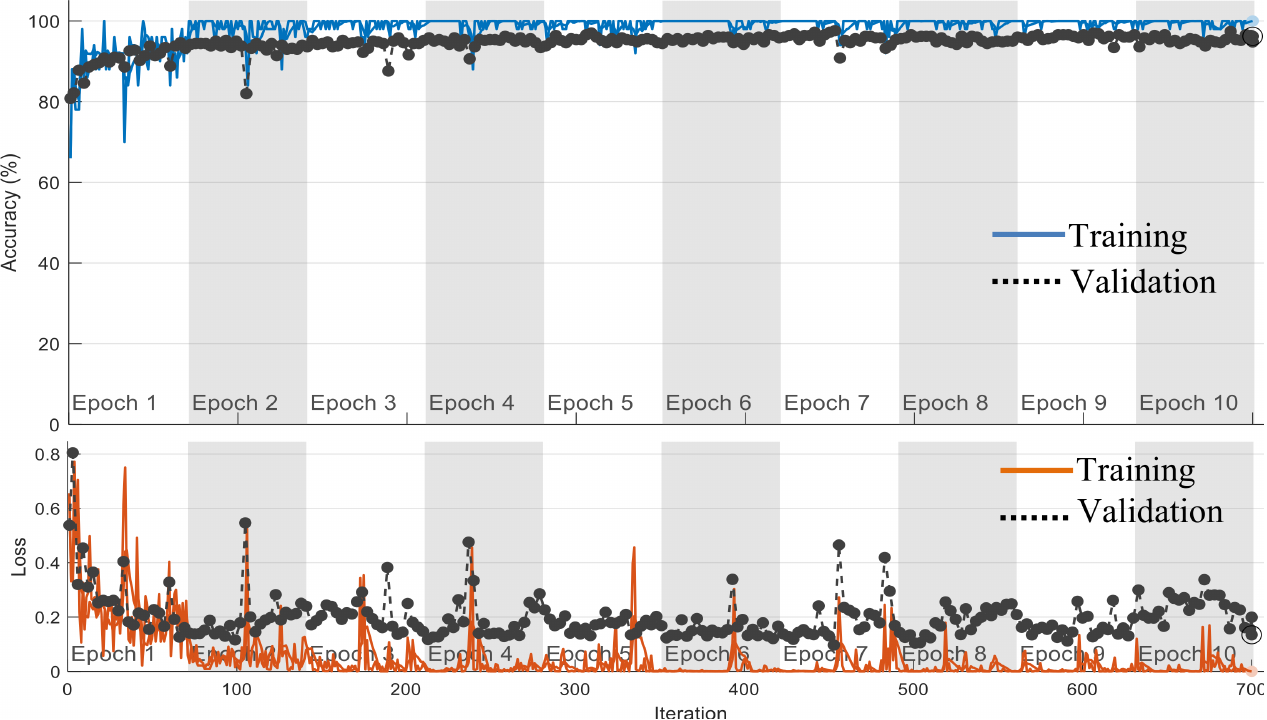} \\
(b)
\end{small}
 \caption{Performance metrics of CNN models: accuracy and loss values of training and validation images using DarkNet50 (\textbf{a}) and ResNet50 (\textbf{b}) models.}
\label{fig4}
\end{figure}

\begin{figure*}[t!]
 \centering
 \includegraphics[width=0.98\textwidth]{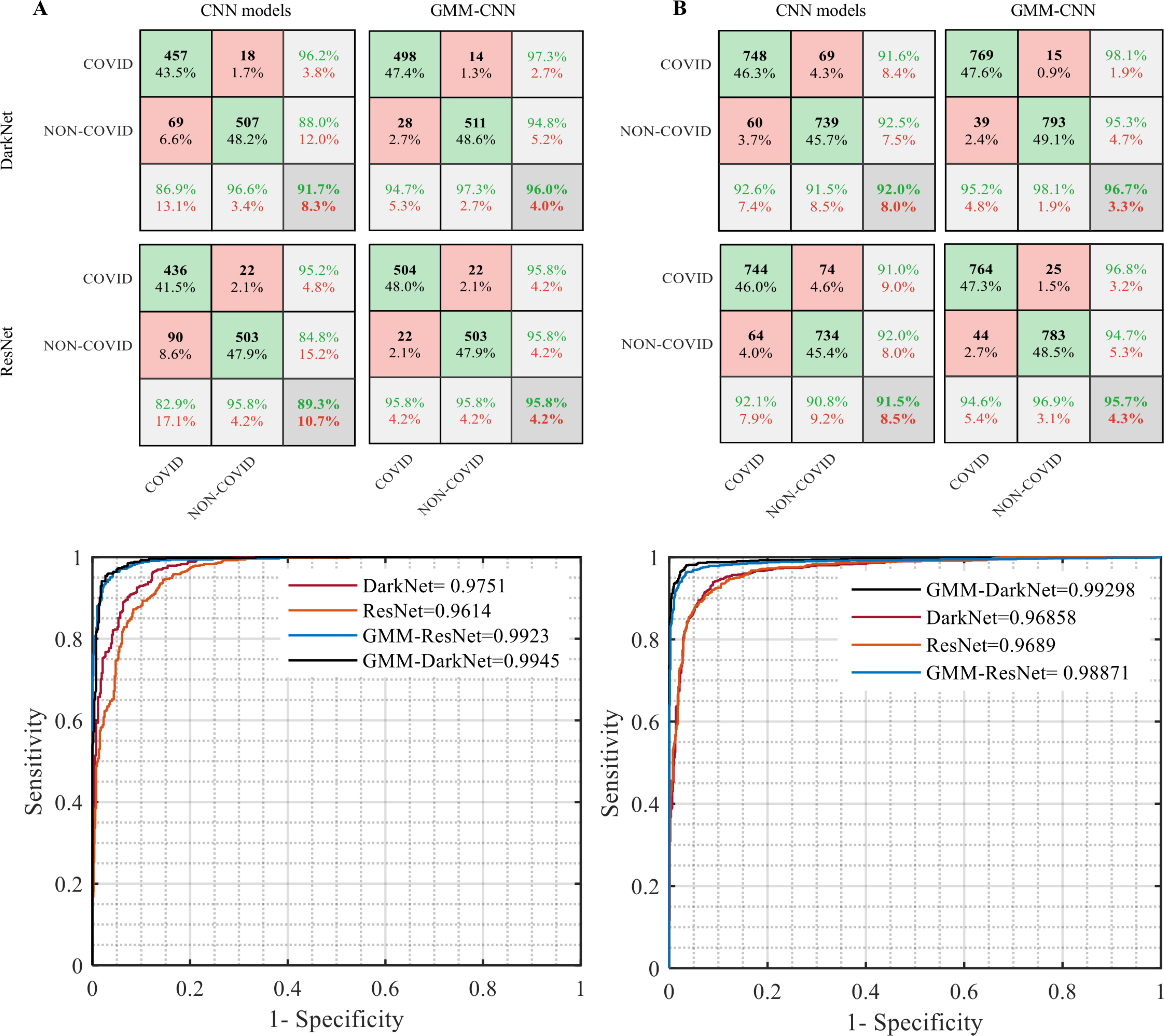}
 \caption{(Top) Confusion matrix obtained on the test (20\%); (bottom) Receiver operating characteristic (ROC)-AUC curve using CT (\textbf{A}) and X-ray images (\textbf{B}).}
 \label{fig5}
\end{figure*}

Next, we compare our GMM-CNN\,+\,RF model against predicting the COVID-19 and NON-COVID-19 classes directly with the DarkNet or ResNet architectures. Fig. \ref{fig4} provides an example of the accuracy and loss function values during training (70\% training and 10\% validation). As can be seen, a growing gap appears between the training and validation loss/accuracy curves. This indicates that the networks are overfitting the training data, as expected. 

Fig. \ref{fig5} compares the performance of our GMM-CNN model with the approach using the CNNs directly, on the test data (20\% of examples). Using the CNN's prediction yields an accuracy of ~89\,--\,91\% for CT (Fig. \ref{fig5}A) and ~91\,--\,92\% for X-ray (Fig. \ref{fig5}A). The accuracy for CT and X-ray is improved by ~5\,--\,6\% when instead using the GMM-CNN features as input to the RF classifier. Likewise, employing our GMM-CNN\,+\,RF method to predict the COVID-19 in CT and X-ray images boosts the AUC-ROC by $\sim$3\%. Table \ref{table2} reports the p-values of Chi-square tests comparing the baseline models (DarKnet and ResNet) to our GMM-CNN\,+\,RF method. As can be seen, our method gives a significantly higher performance (p\,$<$\,0.05), for all combinations of image modality and network architecture. 

In the next experiment, we consider the axial CT and X-ray images and train the RF model on GMM-CNN features obtained from both the DarkNet and ResNet networks. We then compare our GMM-CNN model against running PCA on feature maps ($\mr{PC}\!=\!3$) or using the CNNs directly for prediction. As reported in Table \ref{table3}, using combined GMM-DarkNet and GMM-ResNet features as input to the RF classifier achieves the highest performance in all cases, with an accuracy, AUC, sensitivity and precision of 97.00\%, 99.45\%, 98.10\% and 97.67\%, respectively. In terms of accuracy, our method improves standard CNN classification by 1.9\%\,--\,2.0\%, and PCA by 1.5\%.

In Table \ref{table4}, we compare the results of our method with those reported in recent COVID-19 classification studies. We use accuracy and AUC for this comparison, as these metrics are the most commonly-reported ones in the literature. We find that our results are comparable with the state-of-art found in \cite{78}. Specifically, we obtain a similar accuracy (97\% versus 91.82\,--\,97.48\%) and a higher AUC (99.45\% versus 98.00\,--\,99.00\%) to this previous work which is based on a large dataset (i.e., 10\,--\,15k examples) of X-ray images.

\begin{table}[ht!]
\centering
\caption{Performance (\%) of using DarkNet and ResNet directly, or using DarkNet+ResNet features encoded with PCA or GMM-CNN ($k\!=3$).}\label{table3}
\setlength{\tabcolsep}{2pt}
\renewcommand{\arraystretch}{1.1}
\begin{small}
\begin{tabular}{lccccc}
\toprule
Model & Accuracy & AUC & Sensitivity & Precision \\
\midrule
DarkNet & 95.10& 98.37& 96.32& 95.13\\
ResNet & 94.80& 98.64& 93.43& 95.06 \\
PCA (Dark+Res) & 95.53& 98.56& 96.38& 95.34\\
GMM-CNN (Dark+Res) & 97.00 & 99.45 & 98.10 & 97.67\\
\bottomrule
\end{tabular}
\end{small}
\end{table}

\begin{table}[ht!]
\centering
\caption{Performance (\%) of baselines and our method using CT and/or X-ray scans.
}\label{table4}
\setlength{\tabcolsep}{6pt}
\renewcommand{\arraystretch}{1.1}
\begin{small}
\begin{tabular}{lcccc}
\toprule
Model & Accuracy & AUC & Imaging \\
\midrule
Zhao et al.~\cite{68} & 89.00& 98.00& CT\\
Loey et al.~\cite{69} & 82.91& -- & CT \\
Maghdid et al.~\cite{31} & 94.00\,--\,94.10& --& CT+X-ray \\
Li et al.~\cite{70} & --& 96.00& CT \\
Khan et al.~\cite{78}& 91.82\,--\,97.48 & 98.00--99.00& X-ray \\
\midrule
Ours ($k\!=\!2$) & 96.67& 99.18 & CT+X-ray\\
Ours ($k\!=\!3$) & 97.00 & 99.45 & CT+Xray\\
\bottomrule
\end{tabular}
\end{small}
\end{table}

\section{Discussions}

The assessment for COVID-19 is typically performed with an RT-PCR test, which may be followed by chest CT or X-ray imaging \cite{soedarsono2021management}. Since COVID-19 is a recently discovered disease, this assessment may however vary across different experts and even regions \cite{strully2021regional}. A fast and effective model to detect, predict and screen COVID-19 patients is thus needed for managing the current situation worldwide. AI models based on image analysis provide a noninvasive and effective way to distinguish COVID-19 from other viral pneumonia \cite{71}. Moreover, CT and X-ray images of COVID-19 patients can show abnormal tissues (lesions) that are not detected by early-stage radiography \cite{chamorro2021radiologic}. By learning to detect such anomalies, AI models could improve the screening and prediction COVID-19~\cite{70}. Despite their outstanding performance on applications involving natural images, AI models based on deep learning have had a more limited success in medical image analysis due to the scarcity of annotated data. To alleviate this problem, we proposed a radiomic model, called GMM-CNN, which uses features learned from a deep CNN, and encoded with GMMs, to train a robust classifier based on random forests.

The experiments of this study demonstrated the advantage of our model compared to standard classification using deep CNNs (i.e., DarkNet and ResNet). Specifically, our results showed that a RF classifier trained with the proposed GMM-CNN features provides significant improvements (p\,$<$\,0.05) over CNN-based classification for discriminating between COVID-19 and other viral pneumonia (c.f., Fig. \ref{fig5}). Our experiments also showed that the distribution of features in each CNN feature map could be modeled effectively with very few mixture components ($k\!=\!2,3$), giving rise to a highly compact and discriminative radiomic descriptor. Moreover, we demonstrated the benefit of combining features extracted from both CT and X-ray images, as well as using different CNN architectures. This is made possible by the use of a RF classifier which has a built-in mechanism for selecting relevant features and reduces overfitting with a bagging strategy. Further, the proposed method also offers outstanding performance in comparison to recent approaches for detecting COVID-19 in CT or X-ray images. In particular, it achieves state-of-the-art AUC in the literature. This is due to its ability to combine information from both imaging modalities, without the risk of overfitting like standard CNN classification.

Despite achieving promising results, our work also has limitations that could be addressed in future studies. For instance, our model does not include clinical and demographics information of patients, such as age, sex, treatments and overall survival, which was not available in the dataset. Adding this information as additional inputs to the RF classifier could further improve prediction accuracy and help provide a more personalized diagnosis and treatment for COVID-19. In future work, we plan on extending our study by applying the same proposed radiomics pipeline to additional databases. We also plan to study the impact COVID-19 on survival outcome, by finding texture variations in lung imaging data and analyzing their relationship to survival. Moreover, we will investigate potential links between our radiomic model and other conditions such as pneumonia, bronchitis and tuberculosis, which may occur along COVID-19 in infected patients.

\section{Conclusion}

We proposed a novel radiomic model, GMM-CNN, to predict COVID-19 in CT and X-ray images. This model uses GMM to encode the distribution of features in different layers of a deep CNN  into a compact and discriminating set of descriptors. Our results showed that using the proposed GMM-CNN features as input to a RF classifier outperforms standard CNN classification, yielding an accuracy of 97\% and an AUC over 99\%. This suggests that the proposed model could be used as an important tool for COVID-19 identification. As GMM-CNN features can be acquired from arbitrary images, the usefulness of our approach could also be tested on modalities other than CT and X-ray. 

{\small
\bibliographystyle{IEEEtran}
\bibliography{egbib}

\begin{thebibliography}{10}
\providecommand{\url}[1]{#1}
\csname url@samestyle\endcsname
\providecommand{\newblock}{\relax}
\providecommand{\bibinfo}[2]{#2}
\providecommand{\BIBentrySTDinterwordspacing}{\spaceskip=0pt\relax}
\providecommand{\BIBentryALTinterwordstretchfactor}{4}
\providecommand{\BIBentryALTinterwordspacing}{\spaceskip=\fontdimen2\font plus
\BIBentryALTinterwordstretchfactor\fontdimen3\font minus
  \fontdimen4\font\relax}
\providecommand{\BIBforeignlanguage}[2]{{%
\expandafter\ifx\csname l@#1\endcsname\relax
\typeout{** WARNING: IEEEtran.bst: No hyphenation pattern has been}%
\typeout{** loaded for the language `#1'. Using the pattern for}%
\typeout{** the default language instead.}%
\else
\language=\csname l@#1\endcsname
\fi
#2}}
\providecommand{\BIBdecl}{\relax}
\BIBdecl

\bibitem{1}
C.~S.~G. of~the International \emph{et~al.}, ``The species {S}evere acute
  respiratory syndrome-related coronavirus: classifying {2019-nCoV} and naming
  it {SARS-CoV-2},'' \emph{Nature microbiology}, vol.~5, no.~4, p. 536, 2020.

\bibitem{3}
H.~Shi, X.~Han, N.~Jiang, Y.~Cao, O.~Alwalid, J.~Gu, Y.~Fan, and C.~Zheng,
  ``Radiological findings from 81 patients with {COVID-19} pneumonia in
  {W}uhan, {C}hina: a descriptive study,'' \emph{The Lancet infectious
  diseases}, vol.~20, no.~4, pp. 425--434, 2020.

\bibitem{7}
N.~Zhu, D.~Zhang, W.~Wang, X.~Li, B.~Yang, J.~Song, X.~Zhao, B.~Huang, W.~Shi,
  R.~Lu \emph{et~al.}, ``A novel coronavirus from patients with pneumonia in
  {C}hina, 2019,'' \emph{New England journal of medicine}, 2020.

\bibitem{9}
M.~Chung, A.~Bernheim, X.~Mei, N.~Zhang, M.~Huang, X.~Zeng, J.~Cui, W.~Xu,
  Y.~Yang, Z.~A. Fayad \emph{et~al.}, ``{CT} imaging features of 2019 novel
  coronavirus ({2019-nCoV}),'' \emph{Radiology}, vol. 295, no.~1, pp. 202--207,
  2020.

\bibitem{11}
Y.~Fang, H.~Zhang, J.~Xie, M.~Lin, L.~Ying, P.~Pang, and W.~Ji, ``Sensitivity
  of chest {CT} for {COVID-19}: comparison to {RT-PCR},'' \emph{Radiology},
  vol. 296, no.~2, pp. E115--E117, 2020.

\bibitem{12}
T.~Ai, Z.~Yang, H.~Hou, C.~Zhan, C.~Chen, W.~Lv, Q.~Tao, Z.~Sun, and L.~Xia,
  ``Correlation of chest {CT} and {RT-PCR} testing for coronavirus disease 2019
  ({COVID-19}) in {C}hina: a report of 1014 cases,'' \emph{Radiology}, vol.
  296, no.~2, pp. E32--E40, 2020.

\bibitem{14}
L.~Li, L.~Qin, Z.~Xu, Y.~Yin, X.~Wang, B.~Kong, J.~Bai, Y.~Lu, Z.~Fang, Q.~Song
  \emph{et~al.}, ``Artificial intelligence distinguishes {COVID-19} from
  community acquired pneumonia on chest {CT},'' \emph{Radiology}, 2020.

\bibitem{76}
J.~P. Kanne, B.~P. Little, J.~H. Chung, B.~M. Elicker, and L.~H. Ketai,
  ``Essentials for radiologists on {COVID-19}: an update—radiology scientific
  expert panel,'' 2020.

\bibitem{77}
N.~Sverzellati, C.~J. Ryerson, G.~Milanese, E.~A. Renzoni, A.~Volpi,
  P.~Spagnolo, F.~Bonella, I.~Comelli, P.~Affanni, L.~Veronesi \emph{et~al.},
  ``Chest {X}-ray or {CT} for {COVID-19} pneumonia? {C}omparative study in a
  simulated triage setting,'' \emph{European Respiratory Journal}, 2021.

\bibitem{17}
H.~X. Bai, B.~Hsieh, Z.~Xiong, K.~Halsey, J.~W. Choi, T.~M.~L. Tran, I.~Pan,
  L.-B. Shi, D.-C. Wang, J.~Mei \emph{et~al.}, ``Performance of radiologists in
  differentiating {COVID-19} from {non-COVID-19} viral pneumonia at chest
  {CT},'' \emph{Radiology}, vol. 296, no.~2, pp. E46--E54, 2020.

\bibitem{81}
N.~Paluru, A.~Dayal, H.~B. Jenssen, T.~Sakinis, L.~R. Cenkeramaddi, J.~Prakash,
  and P.~K. Yalavarthy, ``{Anam-Net}: {A}namorphic depth embedding-based
  lightweight {CNN} for segmentation of anomalies in {COVID-19} chest {CT}
  images,'' \emph{IEEE Transactions on Neural Networks and Learning Systems},
  vol.~32, no.~3, pp. 932--946, 2021.

\bibitem{tamal2021integrated}
M.~Tamal, M.~Alshammari, M.~Alabdullah, R.~Hourani, H.~A. Alola, and T.~M.
  Hegazi, ``An integrated framework with machine learning and radiomics for
  accurate and rapid early diagnosis of covid-19 from chest x-ray,''
  \emph{Expert systems with applications}, vol. 180, p. 115152, 2021.

\bibitem{21}
A.~Chaddad, C.~Desrosiers, M.~Toews, and B.~Abdulkarim, ``Predicting survival
  time of lung cancer patients using radiomic analysis,'' \emph{Oncotarget},
  vol.~8, no.~61, p. 104393, 2017.

\bibitem{84}
B.~Ye, X.~Yuan, Z.~Cai, and T.~Lan, ``Severity assessment of {COVID-19} based
  on feature extraction and v-descriptors,'' \emph{IEEE Transactions on
  Industrial Informatics}, vol.~17, no.~11, pp. 7456 -- 7467, 2021.

\bibitem{23}
F.~Shi, L.~Xia, F.~Shan, B.~Song, D.~Wu, Y.~Wei, H.~Yuan, H.~Jiang, Y.~He,
  Y.~Gao \emph{et~al.}, ``Large-scale screening to distinguish between
  {COVID-19} and community-acquired pneumonia using infection size-aware
  classification,'' \emph{Physics in Medicine \& Biology}, vol.~66, no.~6, p.
  065031, 2021.

\bibitem{26}
F.~Shi, J.~Wang, J.~Shi, Z.~Wu, Q.~Wang, Z.~Tang, K.~He, Y.~Shi, and D.~Shen,
  ``Review of artificial intelligence techniques in imaging data acquisition,
  segmentation, and diagnosis for covid-19,'' \emph{IEEE Reviews in Biomedical
  Engineering}, vol.~14, pp. 4--15, 2021.

\bibitem{27}
D.~S. Kermany, M.~Goldbaum, W.~Cai, C.~C. Valentim, H.~Liang, S.~L. Baxter,
  A.~McKeown, G.~Yang, X.~Wu, F.~Yan \emph{et~al.}, ``Identifying medical
  diagnoses and treatable diseases by image-based deep learning,'' \emph{Cell},
  vol. 172, no.~5, pp. 1122--1131, 2018.

\bibitem{85}
A.~Chaddad, L.~Hassan, and C.~Desrosiers, ``Deep cnn models for predicting
  covid-19 in ct and x-ray images,'' \emph{Journal of Medical Imaging}, vol.~8,
  no.~S1, p. 014502, 2021.

\bibitem{bejani2021systematic}
M.~M. Bejani and M.~Ghatee, ``A systematic review on overfitting control in
  shallow and deep neural networks,'' \emph{Artificial Intelligence Review},
  pp. 1--48, 2021.

\bibitem{28}
A.~Chaddad, M.~Toews, C.~Desrosiers, and T.~Niazi, ``Deep radiomic analysis
  based on modeling information flow in convolutional neural networks,''
  \emph{IEEE Access}, vol.~7, pp. 97\,242--97\,252, 2019.

\bibitem{82}
Y.~Oh, S.~Park, and J.~C. Ye, ``Deep learning {COVID-19} features on {CXR}
  using limited training data sets,'' \emph{IEEE Transactions on Medical
  Imaging}, vol.~39, no.~8, pp. 2688--2700, 2020.

\bibitem{30}
S.~Singh, K.~Ho-Shon, S.~Karimi, and L.~Hamey, ``Modality classification and
  concept detection in medical images using deep transfer learning,'' in
  \emph{2018 International Conference on Image and Vision Computing New Zealand
  (IVCNZ)}.\hskip 1em plus 0.5em minus 0.4em\relax IEEE, 2018, pp. 1--9.

\bibitem{39}
A.~Shamsi, H.~Asgharnezhad, S.~S. Jokandan, A.~Khosravi, P.~M. Kebria,
  D.~Nahavandi, S.~Nahavandi, and D.~Srinivasan, ``An uncertainty-aware
  transfer learning-based framework for {COVID-19} diagnosis,'' \emph{IEEE
  Transactions on Neural Networks and Learning Systems}, vol.~32, no.~4, pp.
  1408--1417, 2021.

\bibitem{42}
L.~Fu, Y.~Li, A.~Cheng, P.~Pang, and Z.~Shu, ``A novel machine learning-derived
  radiomic signature of the whole lung differentiates stable from progressive
  {COVID-19} infection: a retrospective cohort study,'' \emph{Journal of
  thoracic imaging}, vol.~35, no.~6, p. 361, 2020.

\bibitem{45}
S.~S. Yip and H.~J. Aerts, ``Applications and limitations of radiomics,''
  \emph{Physics in Medicine \& Biology}, vol.~61, no.~13, p. R150, 2016.

\bibitem{46}
M.~A. Mazurowski, M.~Buda, A.~Saha, and M.~R. Bashir, ``Deep learning in
  radiology: {A}n overview of the concepts and a survey of the state of the art
  with focus on {MRI},'' \emph{Journal of magnetic resonance imaging}, vol.~49,
  no.~4, pp. 939--954, 2019.

\bibitem{huang2021artificial}
S.~Huang, J.~Yang, S.~Fong, and Q.~Zhao, ``Artificial intelligence in the
  diagnosis of covid-19: challenges and perspectives,'' \emph{International
  Journal of Biological Sciences}, vol.~17, no.~6, p. 1581, 2021.

\bibitem{chaddad2018deep}
A.~Chaddad, C.~Desrosiers, and T.~Niazi, ``Deep radiomic analysis of mri
  related to alzheimer’s disease,'' \emph{Ieee Access}, vol.~6, pp.
  58\,213--58\,221, 2018.

\bibitem{47}
A.~Chaddad, M.~Zhang, C.~Desrosiers, and T.~Niazi, ``Deep radiomic features
  from {MRI} scans predict survival outcome of recurrent glioblastoma,'' in
  \emph{International Workshop on Radiomics and Radiogenomics in
  Neuro-oncology}.\hskip 1em plus 0.5em minus 0.4em\relax Springer, 2019, pp.
  36--43.

\bibitem{49}
D.~A. Reynolds, ``Gaussian mixture models.'' \emph{Encyclopedia of biometrics},
  vol. 741, pp. 659--663, 2009.

\bibitem{48}
A.~P. Dempster, N.~M. Laird, and D.~B. Rubin, ``Maximum likelihood from
  incomplete data via the {EM} algorithm,'' \emph{Journal of the Royal
  Statistical Society: Series B (Methodological)}, vol.~39, no.~1, pp. 1--22,
  1977.

\bibitem{51}
Z.~Ma and A.~Leijon, ``Bayesian estimation of beta mixture models with
  variational inference,'' \emph{IEEE Transactions on Pattern Analysis and
  Machine Intelligence}, vol.~33, no.~11, pp. 2160--2173, 2011.

\bibitem{52}
A.~Chaddad, ``Automated feature extraction in brain tumor by magnetic resonance
  imaging using gaussian mixture models,'' \emph{International Journal of
  Biomedical Imaging}, vol. 2015, 2015.

\bibitem{53}
F.~G. Zanjani, S.~Zinger \emph{et~al.}, ``Deep convolutional gaussian mixture
  model for stain-color normalization of histopathological images,'' in
  \emph{International Conference on Medical Image Computing and
  Computer-Assisted Intervention}.\hskip 1em plus 0.5em minus 0.4em\relax
  Springer, 2018, pp. 274--282.

\bibitem{54}
Z.~Liu, Z.~Wu, T.~Li, J.~Li, and C.~Shen, ``{GMM} and {CNN} hybrid method for
  short utterance speaker recognition,'' \emph{IEEE Transactions on Industrial
  informatics}, vol.~14, no.~7, pp. 3244--3252, 2018.

\bibitem{55}
Q.~Guo, F.~Wang, J.~Lei, D.~Tu, and G.~Li, ``Convolutional feature learning and
  hybrid {CNN-HMM} for scene number recognition,'' \emph{Neurocomputing}, vol.
  184, pp. 78--90, 2016.

\bibitem{56}
K.~Sekaran, P.~Chandana, N.~M. Krishna, and S.~Kadry, ``Deep learning
  convolutional neural network ({CNN}) with gaussian mixture model for
  predicting pancreatic cancer,'' \emph{Multimedia Tools and Applications},
  vol.~79, no.~15, pp. 10\,233--10\,247, 2020.

\bibitem{57}
A.~Chaddad, P.~Sargos, and C.~Desrosiers, ``Modeling texture in deep 3d cnn for
  survival analysis,'' \emph{IEEE Journal of Biomedical and Health
  Informatics}, vol.~25, no.~7, pp. 2454--2462, 2021.

\bibitem{58}
G.~Litjens, T.~Kooi, B.~E. Bejnordi, A.~A.~A. Setio, F.~Ciompi, M.~Ghafoorian,
  J.~A. Van Der~Laak, B.~Van~Ginneken, and C.~I. Sanchez, ``A survey on deep
  learning in medical image analysis,'' \emph{Medical image analysis}, vol.~42,
  pp. 60--88, 2017.

\bibitem{59}
K.~He, X.~Zhang, S.~Ren, and J.~Sun, ``Deep residual learning for image
  recognition,'' in \emph{Proceedings of the IEEE conference on computer vision
  and pattern recognition}, 2016, pp. 770--778.

\bibitem{60}
J.~Redmon, ``Darknet: {O}pen source neural networks in c,'' 2013.

\bibitem{62}
S.~A. Taghanaki, A.~Bentaieb, A.~Sharma, S.~K. Zhou, Y.~Zheng, B.~Georgescu,
  P.~Sharma, Z.~Xu, D.~Comaniciu, and G.~Hamarneh, ``Select, attend, and
  transfer: {L}ight, learnable skip connections,'' in \emph{International
  Workshop on Machine Learning in Medical Imaging}.\hskip 1em plus 0.5em minus
  0.4em\relax Springer, 2019, pp. 417--425.

\bibitem{80}
S.~Hussain, A.~Khan, and M.~Zafar, ``Coronavirus disease analysis using chest
  {X}-ray images and a novel deep convolutional neural network,''
  \emph{10.13140/Rg. 2.2. 35868.64646}, no. April, pp. 1--31, 2020.

\bibitem{65}
D.~Dittman, T.~M. Khoshgoftaar, R.~Wald, and A.~Napolitano, ``Random forest:
  {A} reliable tool for patient response prediction,'' in \emph{2011 IEEE
  International Conference on Bioinformatics and Biomedicine Workshops
  (BIBMW)}.\hskip 1em plus 0.5em minus 0.4em\relax IEEE, 2011, pp. 289--296.

\bibitem{66}
R.~Pal, \emph{Predictive modeling of drug sensitivity}.\hskip 1em plus 0.5em
  minus 0.4em\relax Academic Press, 2016.

\bibitem{67}
N.~S. Sumi, M.~A. Islam, and M.~A. Hossain, ``Evaluation and computation of
  diagnostic tests: a simple alternative,'' \emph{Bull Malaysian Math Sci Soc},
  vol.~37, pp. 411--23, 2014.

\bibitem{paluszek2020practical}
M.~Paluszek and S.~Thomas, ``Practical matlab deep learning,'' 2020.

\bibitem{paluszek2016matlab}
------, \emph{MATLAB machine learning}.\hskip 1em plus 0.5em minus 0.4em\relax
  Apress, 2016.

\bibitem{78}
S.~H. Khan, A.~Sohail, and A.~Khan, ``Covid-19 detection in chest x-ray images
  using a new channel boosted cnn,'' \emph{arXiv preprint arXiv:2012.05073},
  2020.

\bibitem{68}
J.~Zhao, Y.~Zhang, X.~He, and P.~Xie, ``{COVID-CT-dataset}: a {CT} scan dataset
  about {COVID-19},'' \emph{arXiv preprint arXiv:2003.13865}, vol. 490, 2020.

\bibitem{69}
M.~Loey, G.~Manogaran, and N.~E.~M. Khalifa, ``A deep transfer learning model
  with classical data augmentation and {CGAN} to detect {COVID-19} from chest
  {CT} radiography digital images,'' \emph{Neural Computing and Applications},
  pp. 1--13, 2020.

\bibitem{31}
H.~S. Maghdid, A.~T. Asaad, K.~Z. Ghafoor, A.~S. Sadiq, S.~Mirjalili, and M.~K.
  Khan, ``Diagnosing {COVID-19} pneumonia from {X}-ray and {CT} images using
  deep learning and transfer learning algorithms,'' in \emph{Multimodal Image
  Exploitation and Learning 2021}, vol. 11734.\hskip 1em plus 0.5em minus
  0.4em\relax International Society for Optics and Photonics, 2021, p. 117340E.

\bibitem{70}
L.~Li, L.~Qin, Z.~Xu, Y.~Yin, X.~Wang, B.~Kong, J.~Bai, Y.~Lu, Z.~Fang, Q.~Song
  \emph{et~al.}, ``Artificial intelligence distinguishes {COVID-19} from
  community acquired pneumonia on chest {CT},'' \emph{Radiology}, 2020.

\bibitem{soedarsono2021management}
S.~Soedarsono, A.~Febriani, H.~Hasan, and A.~Widyoningroem, ``Management of
  severe covid-19 patient with negative rt-pcr for sars-cov-2: Role of
  clinical, radiological, and serological diagnosis,'' \emph{Radiology Case
  Reports}, vol.~16, no.~6, pp. 1405--1409, 2021.

\bibitem{strully2021regional}
K.~Strully, T.-C. Yang, and H.~Liu, ``Regional variation in covid-19
  disparities: connections with immigrant and latinx communities in us
  counties,'' \emph{Annals of Epidemiology}, vol.~53, pp. 56--62, 2021.

\bibitem{71}
Y.~Mohamadou, A.~Halidou, and P.~T. Kapen, ``A review of mathematical modeling,
  artificial intelligence and datasets used in the study, prediction and
  management of {COVID-19},'' \emph{Applied Intelligence}, vol.~50, no.~11, pp.
  3913--3925, 2020.

\bibitem{chamorro2021radiologic}
E.~M. Chamorro, A.~D. Tascon, L.~I. Sanz, S.~O. Velez, and S.~B. Nacenta,
  ``Radiologic diagnosis of patients with covid-19,'' \emph{Radiologia (English
  Edition)}, vol.~63, no.~1, pp. 56--73, 2021.

\end{thebibliography}
}
\end{document}